\documentclass[graybox]{svmult} 
 
\usepackage{mathptmx}       
\usepackage{helvet}         
\usepackage{courier}        
\usepackage{type1cm}        
\usepackage{makeidx}         
\usepackage{graphicx}        
\usepackage{multicol}        
\usepackage[bottom]{footmisc}
 
\makeindex             
 
\begin{document} 
 
\title*{Period doubling in {\it Kepler} RR Lyrae stars} 
\author{R. Szab\'o, Z. Koll\'ath, L. Moln\'ar, K. Kolenberg, D.~W. Kurtz and 
WG\#13 members} 
\institute{R. Szab\'o, Z. Koll\'ath, L. Moln\'ar \at Konkoly Observatory of the 
Hungarian Academy of  
Sciences, H-1121 Budapest, Konkoly Thege Mikl\'os \'ut 15-17., Hungary, 
\email{rszabo@konkoly.hu} 
\and K. Kolenberg \at Harvard-Smithsonian Center for Astrophysics, 60 Garden St.,  
Cambridge MA 02138 USA \\ Instituut voor Sterrenkunde, Celestijnenlaan 200D, 3001 
Heverlee, Belgium 
\and D. Kurtz \at Jeremiah Horrocks Institute, University of Central Lancashire, 
Preston PR1 2HE, UK} 
%
%
\maketitle 
 
\abstract*{The origin of the conspicuous amplitude and phase modulation of RR\,Lyr 
star pulsation -- known as the Blazhko effect -- is still a mystery more than 100 
years after its discovery. With the help of the {\it Kepler} space telescope we 
have revealed a new and unexpected phenomenon: period doubling in RR\,Lyrae itself 
-- the eponym and prototype of its class -- as well as in other {\it Kepler} 
Blazhko RR\,Lyr stars. We have found that period doubling is directly connected to 
the Blazhko modulation. Furthermore, with hydrodynamic model calculations we have 
succeeded in reproducing the period doubling and proved that the cause of this 
effect is a high order resonance (9:2) between the fundamental mode and the 
$\rm{9}^{th}$ radial overtone, which is a strange mode. We discuss the 
implications of these recent findings on our understanding of the century-old 
Blazhko problem.} 
 
\abstract{The origin of the conspicuous amplitude and phase modulation of the 
RR\,Lyrae pulsation -- known as the Blazhko effect -- is still a mystery after 
more than 100 years of its discovery. With the help of the Kepler space telescope 
we have revealed a new and unexpected phenomenon: period doubling in RR\,Lyr -- 
the eponym and prototype of its class -- as well as in other Kepler Blazhko 
RR\,Lyrae stars. We have found that period doubling is directly connected to the 
Blazhko modulation. Furthermore, with hydrodynamic model calculations we have 
succeeded in reproducing the period doubling and proved that the root cause of 
this effect is a high order resonance (9:2) between the fundamental mode and the 
$\rm{9}^{th}$ radial overtone, which is a strange mode. We discuss the 
implications of these recent findings on our understanding of the century-old 
Blazhko problem.} 
 
\section{Period doubling in pulsating variable stars}\label{sec:1} 
 
\index{period doubling}Period doubling (PD) bifurcation is a well-known dynamical 
effect. In the parlance of dynamical systems a new limit cycle emerges from an 
existing limit cycle with a period twice as long as the old one. In the case of a 
pulsating star we observe alternating cycles in the time domain, in the frequency 
domain PD manifests itself as half-integer frequencies (the f/2 subharmonic and 
its odd integer multiples). In stellar astrophysics the heyday of period doubling 
occurred more than two decades ago when it was discovered in one-dimensional 
hydrodynamic models of \index{Cepheid}Cepheids \cite{MB90}, \cite{MB91}, 
\cite{BM92} and Type II Cepheids \cite{BK87}, \cite{KB88} by J.~R. Buchler and his 
collaborators. The cause of PD was found to be a low-order resonance between the 
fundamental mode and a low radial overtone. In the hydrodynamic model calculations 
mentioned usually the 3:2 or the 5:2 resonances were acting \cite{MB90}. 
One of the main reasons to study the phenomenon is that a star can go from regular 
pulsations to \index{chaos}chaos through a PD bifurcation cascade (known as a 
Feigenbaum cascade). Buchler et al. \cite{BKSM96} proved that the characteristic 
light variations of \index{RV Tauri}RV Tauri stars, which show alternating deep 
and shallow minima, can be interpreted as a result of deterministic 
\index{chaos}chaos of low dimension. 
 
Even a \index{Mira}Mira star, \index{R Cyg}(R~Cyg), was observed to show this 
phenomenon \cite{KSz02}, and quite recently \index{BL Her}BL~Herculis stars with 
PD were discovered in the OGLE-III data \cite{Smolec11}, confirming earlier 
theoretical predictions. \index{RR\,Lyrae} RR\,Lyr stars, on the other hand, have 
been thought to pulsate quite regularly without low order resonances. This belief 
was based on decades-long observations and hydrodynamic models, with the only 
disturbing fact being the perplexing presence of amplitude and phase modulation 
(known as the \index{Blazhko effect} Blazhko effect) in an increasing number of 
these stars. Therefore, the discovery of PD in the {\it Kepler} data was a 
surprise, forcing us to overhaul existing models and theories of RR\,Lyr 
pulsation. 
 
\begin{figure}[b] 
\includegraphics[width=6cm,angle=270]{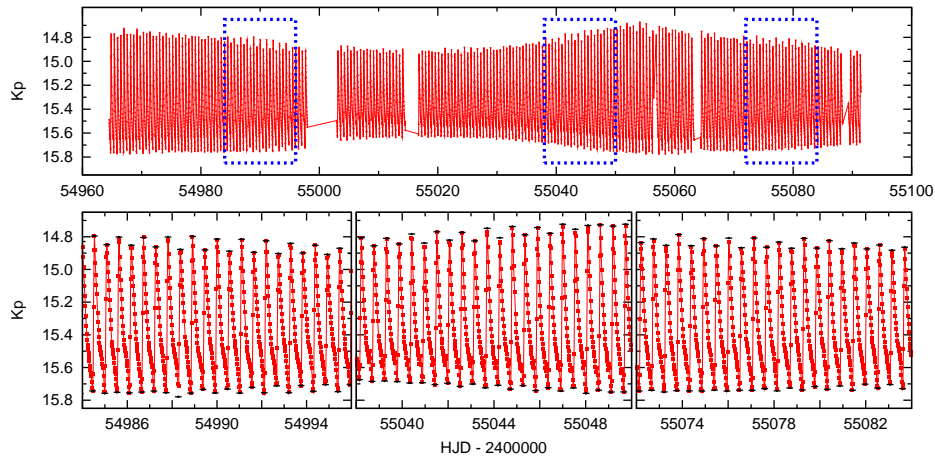} 
\caption{{\bf Top}: A {\it Kepler} light curve of the Blazhko star V808\,Cygni 
(KIC\,4484128) in quarters Q1 and Q2 showing 133\,d of observations. Gaps in the 
data are due to safe mode events of the spacecraft and planned data download 
periods. The Blazhko cycle is around 90\,d. {\bf Bottom}: Magnification of three 
12-d sections of the light curve highlighted by rectangles in the upper panel 
showing the characteristic period doubling behaviour. Polynomial fitting of the 
maxima and minima are also plotted for better visibility.} 
\label{fig:1} 
\end{figure} 
 
\section{Discovery of period doubling in RR\,Lyrae stars}\label{sec:2} 
 
{\it Kepler} is a NASA Discovery mission to find Earth-like planets in the 
habitable zones of solar-like stars using the transit method \cite{Borucki}. It 
provides incredibly high-precision, quasi-continuous observations of a 
115\,$\rm{deg}^{2}$ swath of the sky. The currently known sample of RR\,Lyr stars 
in the {\it Kepler} field consists of some 40 members, the majority of which are 
RRab stars, and half of which show Blazhko modulation. The field contains 
RR\,Lyrae itself, and despite of its brightness, hence heavy saturation on the 
CCD, we managed to get extremely precise photometry for this important target with 
{\it Kepler} \cite{Szabo10}. 
 
We found PD in the first release of data for \index{RR\,Lyrae} RR~Lyrae itself 
(KIC\,7198959) \cite{Kol10}, then subsequently in two other {\it Kepler} Blazhko 
RR\,Lyr stars: \index{V808 Cyg}V808\,Cyg (Fig.\ref{fig:1}) and \index{V355 Lyr}V355\,Lyr 
(KIC\,4484128 and KIC\,7505345, respectively), both of which are much fainter than 
RR\,Lyrae itself \cite{Szabo10}. The strength of the PD is variable; at the 
strongest phase the difference between subsequent maxima can be as large as 0.1 
magnitude in RR\,Lyrae. We found that PD is stronger in certain Blazhko phases. 
There are hints in another four modulated stars of half-integer frequencies 
\cite{Szabo10}, which means that at least half of the modulated RRab stars show PD 
as well. After monitoring our RR\,Lyr star sample with {\it Kepler} for years we 
expect to see PD in more stars. 
 
Interestingly, non-Blazhko stars do not show PD down to the precision of the {\it 
Kepler} measurements \cite{Szabo10}, \cite{Nemec11}. 
Despite close monitoring of RR\,Lyr stars it was not possible to 
detect PD previously, partly because the consecutive cycles rarely can be observed 
from one geographical location, while the usually low amplitude of the phenomenon 
and its non-stationary nature also add to the difficulties. 
 
\section{Period doubling and the Blazhko effect}\label{sec:3} 
 
The PD phenomenon is intimately connected to the Blazhko cycle. Therefore by 
studying it, we may gain new insights into the intricacies of the Blazhko effect. 
Importantly, we succeeded in reproducing PD in hydrodynamic models \cite{Szabo10}, 
and unambiguously traced its cause back to a 9:2 resonance between the fundamental 
mode and the $\rm{9}^{th}$ overtone, which is a \index{strange mode}strange mode 
\cite{Kollath11}, \cite{Molnar11}. 
 
Based on resonant \index{amplitude equations}amplitude equations accounting for 
the 9:2 resonance between the fundamental mode and the $\rm{9}^{th}$ (strange) 
overtone, Buchler \& Koll\'ath \cite{BK11} have found that this resonance may give 
rise not to only period doubled solutions, but irregularly (chaotic) modulated 
solutions as well. This is important, since recent observations proved that the 
Blazhko effect is not a clockwork precision process: both long-term and cycle-to-
cycle variations are frequently found in the modulation (see, e.g., 
\cite{Guggenberger11}). Further investigations should clarify whether these 
amplitude equations can describe state-of-the-art hydrodynamic model calculations 
and, ultimately, real RR\,Lyr stars. If that turns out to be the case, then this 
elegant prediction may be the long-sought explanation of the mysterious Blazhko 
effect. Additional complicating effects can also contribute to the whole picture, 
like resonances involving nonradial modes \cite{ND01}, three-mode resonances 
\cite{Molnar11} and magneto-hydrodynamic dynamo-like processes \cite{Stothers06}. 
Beyond doubt, the discovery of period doubling in the {\it Kepler} data has opened 
a whole new avenue in the research of RR\,Lyr stars. 
 
\begin{acknowledgement} 
We gratefully acknowledge the entire Kepler team, whose outstanding efforts have 
made these results possible. This project has been supported by the `Lend\"ulet' 
program, the Hungarian OTKA grants K83790 
and MB08C 81013, the European Community's Seventh Framework Programme (FP7/2007-2013) 
under grant agreement no. 269194. R. Szab\'o was supported by the J\'anos Bolyai 
Research Scholarship of the Hungarian Academy of Sciences. 
\end{acknowledgement}

\end{document}